\documentclass[10pt,preprint2]{aastex}
\newcommand{\src}{\objectname[3C]{3C~147}}
\newcommand{\kms}{\mbox{km~s$^{-1}$}}
\newcommand{\mjybm}{\mbox{mJy~beam${}^{-1}$}}

\received{2008 June~13}
\revised{2009 February~24}
\accepted{2009 March~1}
\paperid{283526}

\shorttitle{3C~147 and Small-Scale Galactic \ion{H}{1} Structure}
\shortauthors{Lazio et al.}

\begin{document}
\title{Spatial Variations in Galactic \ion{H}{1} Structure on
	AU-Scales Toward 3C~147 Observed with the Very Long Baseline Array}

\author{T.~Joseph~W.~Lazio}
\affil{Naval Research Laboratory, 4555 Overlook Avenue \hbox{SW},
	Washington, DC 20375-5351}
\email{Joseph.Lazio@nrl.navy.mil}

\author{C.~L.~Brogan}
\affil{National Radio Astronomy Observatory, 520 Edgemont Road,
	Charlottesville, VA 22903-2475}

\author{W.~M.~Goss}
\affil{National Radio Astronomy Observatory, P.~O.~Box~O, 1003 
	Lopezville Road, Socorro, NM 87801}

\and

\author{Sne{\v z}ana Stanimirovi{\'c}}
\affil{Department of Astronomy, University of Wisconsin, Madison, WI 53706}

\begin{abstract}
\noindent
This paper reports dual-epoch, Very Long Baseline Array observations
of \ion{H}{1} absorption toward \src.  One of these epochs (2005)
represents new observations while one (1998) represents the
reprocessing of previous observations to obtain higher signal-to-noise
results.  Significant \ion{H}{1} opacity and column density
variations, both spatially and temporally, are observed with typical
variations at the level of $\Delta\tau \approx 0.20$ and in some cases
as large as $\Delta\tau \approx 0.70$, corresponding to column density
fluctuations of order $5 \times 10^{19}$~cm${}^{-2}$ for an assumed
50~K spin temperature.  The typical angular scale is 15~mas; while the
distance to the absorbing gas is highly uncertain, the equivalent
linear scale is likely to be about~10~\hbox{AU}.  Approximately 10\%
of the face of the source is covered by these opacity variations,
probably implying a volume filling factor for the small-scale
absorbing gas of no more than about~1\%.  Comparing our results with
earlier results toward \objectname[3C]{3C~138}
(\citeauthor{bzlgdf05}), we find numerous similarities, and we
conclude that small-scale absorbing gas is a ubiquitious phenomenon,
albeit with a low probability of intercept on any given line of sight.
Further, we compare the volumes sampled by the line of sight through
the Galaxy between our two epochs and conclude that, on the basis of
the motion of the Sun alone, these two volumes are likely to be
substantially different.  In order to place more significant
constraints on the various models for the origin of these small-scale
structures, more frequent sampling is required in any future
observations.
\end{abstract}


\section{Introduction}\label{sec:intro}

Beginning with a two-antenna very long baseline interferometric (VLBI)
observation of \src\ by \cite{dwr76}, a variety of \ion{H}{1}
absorption studies over the past three decades have found AU-scale
optical depth variations in the Galactic interstellar medium (ISM).
The initial detections were confirmed by \cite{dgrbkm89}, and the
first images of the small-scale \ion{H}{1} in the direction of
\objectname[3C]{3C~138} and \src\ were made by \cite{ddg96} using
\hbox{MERLIN}.  \cite{fgdt98} and \cite{fg01} used the Very Long
Baseline Array (VLBA) to improve the resolution toward a number of
sources to approximately 20~mas ($\sim 10$~AU).  Significant
variations were detected in the direction of \objectname[3C]{3C~138}
and \src, while no significant variations in \ion{H}{1} opacity were
found in the direction of five other compact radio sources.

An independent means of probing small-scale neutral structures is
multi-epoch \ion{H}{1} absorption measurements of high proper motion
pulsars \citep{fwcm94,jkww03,swhdg03}. While early pulsar observations
suggested that small-scale structure might be ubiquitous, more recent
observations suggest that it could be more sporadic.  A significant
advantage of VLBI observations is that they provide 2-D images of the
opacity variations, rather than 1-D samples as in the case of pulsars
observations.

\cite{bzlgdf05} revisited the observations of \objectname[3C]{3C~138},
by re-analyzing the 1995 VLBA observations \citep{fgdt98} and by
obtaining two new epochs of observations (1999 and~2002).  They
confirmed the initial results of \cite{fg01}, that there are
small-scale opacity changes along the line of sight to
\objectname[3C]{3C~138} at the level of $\Delta\tau_{\mathrm{max}}
= 0.50 \pm 0.05$, with typical sizes of roughly 50~mas ($\sim 25$~AU).
However, with multiple epochs and improvements in data analysis
techniques (yielding an increase of a factor of~5 in the sensitivity
of the 1995 epoch), they reached a number of additional significant
conclusions:
\begin{enumerate}
\item They found clear evidence for temporal variations in the
\ion{H}{1} opacity over the seven-year time span of the three epochs,
consistent with structures moving across the line of sight at
velocities of a few tens of kilometers per second, though the
infrequent sampling in time means that they could not determine
whether these structures were persistent.

\item They found no evidence for a drop in the \ion{H}{1} spin
temperature, as would be evidenced by a narrowing of line widths at
small scales compared to single dish measurements.  In turn, a
constant \ion{H}{1} spin temperature implies that the small-scale
opacity variations are due to density enhancements, although these
enhancements would necessarily be extremely over-pressured relative to
the mean interstellar pressure, far from equilibrium, and likely of
relatively short duration.

\item For the first time they determined that the plane of sky covering
fraction of the small-scale \ion{H}{1} gas is roughly 10\%.  In turn,
this small covering fraction suggests that the volume filling factor of such
gas, within the cold neutral medium, is quite low ($\lesssim 1$\%), in
agreement with HST observations of high-pressure gas in the ISM
\citep{jt01,J04}.

\item They simulated pulsar observations that have been used to search
for \ion{H}{1} opacity variations and showed that the existing pulsar
observations have generally been too sparsely sampled (in time) to be
useful in studying the details of small-scale \ion{H}{1} opacity variations.
\end{enumerate}

While the multi-epoch study of \cite{bzlgdf05} represented a
substantial improvement, nonetheless their conclusions rested on
observations of only one line of sight.  In light of this sample of
one, their conclusions might seem rather audacious, particularly given
the larger sample observed by \cite{fgdt98} and \cite{fg01}, in which
most of the objects did not show variations in the \ion{H}{1}
absorption.  The \objectname[3C]{3C~138} study has shown that the key
to a successful small scale \ion{H}{1} study is a background source
with both high surface brightness ($\gtrsim 60$~\mjybm) and large
angular extent ($> 100$~mas).  The quasar \src\ is one of the
few sources that shares these characteristics with
\objectname[3C]{3C~138}.  This paper presents dual-epoch observations
of \src\ that were designed specifically to confront the conclusions of
\cite{bzlgdf05} with a second line of sight.
Section~\ref{sec:observe} of this paper describes the observations,
focussing on the new observations acquired for the second epoch,
\S\ref{sec:results} discusses the results, and
\S\ref{sec:conclude} presents our conclusions and recommendations for
future work.

\section{Observations}\label{sec:observe}

We have observed the Galactic \ion{H}{1} absorption (near~$-10\,\kms$)
toward the quasar \src\ at two epochs.  Epoch~I was 1998 October~22,
and the results from those observations have been published previously
by \cite{fg01}.  Epoch~II consists of new data observed on~2005
August~21.  Table~\ref{tab:log} summarizes the basic observing
parameters for the two epochs.

\begin{deluxetable}{lccc}
\tablewidth{0pc}
\tablecaption{Observational Log\label{tab:log}}
\tabletypesize{\footnotesize}
\tablehead{
 \colhead{Parameter}             & \cite{fg01}     & \colhead{Epoch~I} & \colhead{Epoch~II}}
\startdata
Date                             & 1998 October 22 & 1998 October~22   & 2005 August~21 \\
Number of IFs                    & 4                 & 4               & 4 \\
Bandwidth per IF (MHz)           & 0.5               & 0.5             & 0.5 \\
Spectral channels                & 256               & 256             & 512 \\
Channel separation (\kms)        & 0.41              & 0.41            & 0.21 \\
Velocity resolution (\kms)\tablenotemark{a}   & 0.4               & 0.41             & 0.21 \\
\textsc{clean} beam (mas)\tablenotemark{b}    & 5 $\times$ 4      & 8.2 $\times$ 5.6 & 7.6 $\times$ 7.1 \\
Continuum Peak (Jy~beam${}^{-1}$)\tablenotemark{c} & 2.31         & 1.681          & 1.801 \\
Continuum rms noise (\mjybm)\tablenotemark{c}     & 6.5           & 1.7              & 1.1 \\
Spectral line rms noise (\mjybm)\tablenotemark{c} & 7.6           & 5.0               & 3.5 \\
\cutinhead{General Parameters for \src}
Position (equatorial, J2000) &  $05^{\rm h}42^{\rm m}36\fs13788$ & $+49\arcdeg 07\arcmin 51\farcs2335$ \\
Position (Galactic, longitude \& latitude) & $+161.69\arcdeg$     &  $+10.30\arcdeg$ \\ 
Redshift  & 0.545 & \\
\enddata
\tablenotetext{a} {After Hanning smoothing during imaging process.}
\tablenotetext{b} {The \textsc{clean} beam before convolution to~10~mas. 
All subsequent values are for the convolved 10~mas resolution images.}
\tablenotetext{c} {For the image after it has been convolved to~10~mas
resolution.}
\tablecomments{The values listed under the \cite{fg01} column are
for the original analysis.  The values listed under the Epoch~I
column are for this analysis, after the reprocessing of the data as
described in the text.}
\end{deluxetable}

For both epochs the data were obtained using the 10 antennas of the
Very Long Baseline Array combined with the Very Large Array with its
27 antennas operating in a phased-array mode.  For the 2005 epoch,
the Green Bank Telescope was also used.  The observing duration was
12~hr for the 1998 epoch and 16~hr for the 2005 epoch, including time
spent on calibration sources. The proximity of the VLA to the VLBA
antenna at Pie Town, New Mexico, significantly increased our
sensitivity to large-scale structures.
Four separate spectral windows or intermediate frequency bands (IFs)
were used, with one IF centered on the absorption line (at an
approximate LSR velocity of~$-10\,\kms$) and three IFs separated by at
least 100~\kms\ in velocity in order to sample the 21~cm continuum
emission.  For the 1998 epoch, the data were correlated with velocity
channels of~0.4~\kms, with a bandwidth of~500~kHz per IF
over~256~spectral channels; for the 2005 epoch, improvements in the
correlator allowed the number of spectral channels per IF to be
increased to~512, with a concomitant improvement in the velocity
resolution to~0.2~\kms.

Broadly similar data reduction procedures were used for the two
epochs.  For the 2005 epoch, the data were calibrated for the frequency
dependence of the bandpass using observations of
\objectname[3C]{3C~48} and amplitude calibrated using system
temperatures measured at the individual antennas.  The most
significant difference in the calibration is that for the 2005 epoch,
we attempted to  phase-reference the observations to the compact source
\objectname[IVS]{IVS~B0532$+$506}, separated by~1\fdg3 from \src.

Our initial motivation for this change in procedure is that \src\ has
a sufficiently complex structure that fringe-fitting assuming a
point-source model could yield erroneous residual delay and rate
solutions.  In practice, phase referencing did not prove useful.  The
phase-referencing cycle time was short enough that latency in the VLA
system often resulted in the VLA acquiring no data.  The most
significant difficulty, however, was that only one of the epochs was
phase-referenced.  There was an apparent offset in the core position
between the two epochs (with a magnitude of a fraction of the
synthesized beam width or a few milliarcseconds) that biased any
attempt to compare results from the two epochs (e.g., comparing the
integrated line profiles).  Consequently, we did not make use of the
phase-referenced data for constructing the \ion{H}{1} line profile or
opacity images.  One obvious impact on our results is that the
sensitivity of the 2005 epoch \ion{H}{1} line data is less than it
could have otherwise been due to the phase-referencing cycling between
\src\ and \objectname[IVS]{IVS~B0532+506}.

Two of the three continuum IFs were then averaged together and several
iterations of hybrid imaging (iterative imaging and self-calibration)
were performed.  After the final iteration of self-calibration, the
phase and amplitude solutions were applied to the IF containing the
\ion{H}{1} line.  The line-free velocity channels in this IF were
averaged together to produce a continuum data set, which underwent a
final round of hybrid imaging, the solutions from which were applied
to the velocity channels containing the line.  Finally, the continuum
emission was subtracted from the velocity channels containing the
\ion{H}{1} line and the resulting line data set was imaged.

We also reprocessed the observations of \cite{fg01} in a similar
fashion.  A significant difference from the original analysis of
\cite{fg01} is that we used the continuum image from the 2005 epoch as an
initial model for fringe fitting the 1998 epoch data (for which no phase
referencing was performed).  The combination of a better initial model
and improvements in the imaging software and analysis procedures led
to a substantial improvement in the reprocessed 1998 epoch data.  The
noise in the 1998 epoch continuum image has improved by nearly an order
of magnitude, and the improvement in the spectral line images is a
factor of a few.  As was the case for \objectname[3C]{3C~138}, the
original analysis found a significantly higher peak brightness than we
do, by a similar factor ($\approx 30$\%).  Like \cite{bzlgdf05}, we
attribute this difference to the use of a point source model by
\cite{fg01} in the original fringe fitting along with other details of
the subsequent imaging and self-calibration.

Following the procedure of \cite{fg01}, both the continuum images and
continuum-subtracted line cubes were convolved to~10~mas resolution.
The $u$-$v$ coverages for the visibility data from the two epochs were
similar, producing images with angular resolutions of approximately
7~mas (Table~\ref{tab:log}).  The convolution of the continuum images and
continuum-subtracted line cubes is an attempt to minimize any effects
of modest differences in the $u$-$v$ coverage between the epochs.  The
second epoch data were also smoothed in velocity so that their
velocity resolution matched that of the first epoch.

An optical depth cube, calculated as $\tau_{H\,I}(\alpha, \delta, v) =
-\ln[1 - I_{\mathrm{line}}(\alpha, \delta, v)/I_{\mathrm{cont}}(\alpha,
\delta)]$, where $I_{\mathrm{line}}$ and $I_{\mathrm{cont}}$ are,
respectively, the images from formed from line and line-free channels.
Because the signal-to-noise ratio in the optical depth images is low
where the continuum emission is weak, the optical depth images were
blanked where the continuum emission was less than 5\% of the peak
emission.

\section{Results}\label{sec:results}

\subsection{21~cm Continuum}\label{sec:cont}

\begin{figure}[tb]
\epsscale{0.95}
\plotone{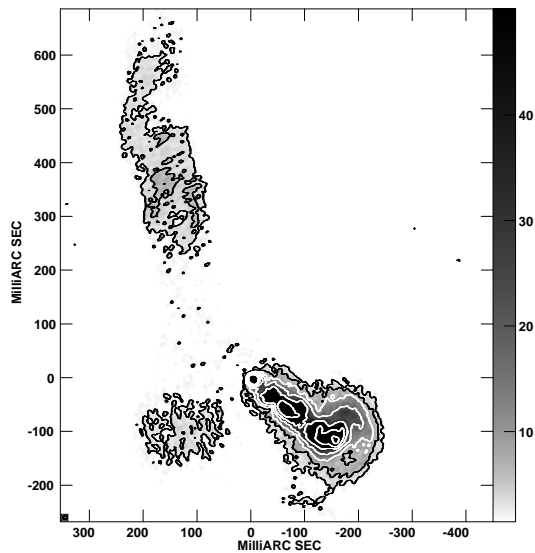}
\vspace*{-3ex}
\caption[]{The Epoch~II (2005) 21~cm continuum image of
\src\ obtained with the VLBA and phased \hbox{VLA}.  The
\textsc{clean} beam is 7.6~mas $\times$ 7.1~mas at a position angle
of~$-15\arcdeg$.  The rms noise level is 0.5~\mjybm, and the contours
are set at 0.5~\mjybm\ $\times$ $-3$, 5, 7.07, 10, 14.1, 20, $\ldots$.
The gray scale is linear between~1.5 and~500~\mjybm.  This image shows
the source at the full resolution; for subsequent analysis, the image
was convolved to~10~mas resolution.  The origin is at (J2000) 
$05^{\mathrm{h}}\,42^{\mathrm{m}}\,36\fs1379$~$+49\arcdeg\,51\arcmin\,07\farcs234$.}
\label{fig:cont}
\end{figure}

Figure~\ref{fig:cont} presents the 21~cm continuum image of \src\ from
the new observations of the 2005 epoch.  There is good qualitative
agreement between our image and previously published images at
comparable wavelengths \citep[18--20~cm,][]{rw80,zac+91,pwx+95,fg01}.
The source displays its well-known core-jet structure, with the jet
extending some 200~mas to the southwest before bending to the north.
Also prominent is diffuse emission to the east of the core, extending
to the north, first noticed by \cite{zac+91}.  The resolution of our
observations is not high enough to resolve the northeast extension
from the core found by \cite{rw80}.

We assessed the continuum images from the two epochs for variability.
Creating a difference image between the two epochs, we find that any
variability in the source is below the 20~\mjybm\ level.  Even if the
source is variable at this level, as for the \cite{bzlgdf05} analysis,
variability will not impact our optical depth calculations, because
(1)~the continuum appropriate for each epoch was used in the optical
depth calculations, (2)~amplitude self-calibration solutions were
never transferred between the epochs, and (3)~the intrinsic continuum
morphology at~10~mas resolution does not appear to change from epoch
to epoch.

The flux density in our image is 18~Jy.  The VLA Calibrator Manual
lists of flux density of~22.5~Jy, indicating that our observations
have recovered 80\% of the source's total flux density.

\subsection{\ion{H}{1} Line Profile}\label{sec:line}

\begin{figure}[bt]
\epsscale{0.95}
\plotone{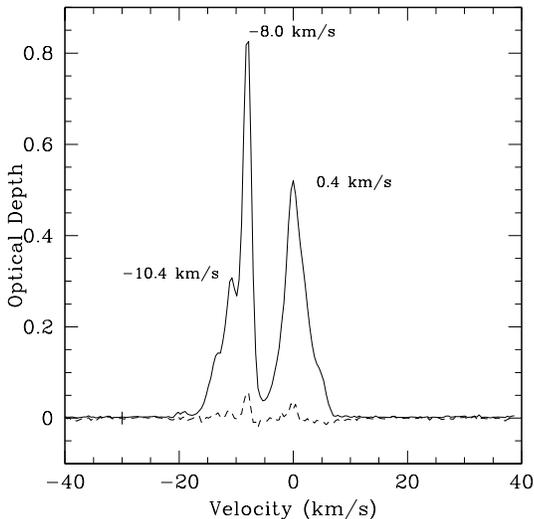}
\vspace*{-3ex}
\caption[]{The solid line shows the average \ion{H}{1} optical depth
profile toward \src\ from Epoch~II (2005).  The dotted line shows the
difference between the line profiles from the two epochs.  Also marked
are the three significant velocity components at which further optical
depth analysis is performed (viz.\ Table~\ref{tab:profile}).}
\label{fig:profile}
\end{figure}

Figure~\ref{fig:profile} shows the average optical depth profile from
the new observation of~2005.  Even with its relatively high latitude
($b = +10\arcdeg$), the profile is complex, making it similar to
\objectname[3C]{3C~138} \citep{bzlgdf05}.  Our profile is in good
agreement with previously published profiles \citep{ksg85,fg01}, and
the difference between the line profiles from the two epochs shows
only modest variations.  Like \cite{fg01}, we shall restrict our
attention to the three most prominent velocity components, those
with approximate central velocities of~$-10.4$, $-8.0$, and~0.4~\kms\
(see below).

\cite{fg01} have discussed the difficulties with assessing the
distance to the absorbing gas.  Under the simple assumption that all
\ion{H}{1} gas is confined to a 100~pc thick layer, the absorbing gas
must be within~0.6~kpc.  However, the kinematic distance to the gas
causing the $-8\,\kms$ absorption is uncertain, with distances as
large as 1.1~kpc allowed.  As a nominal value, we adopt the conversion
that our resolution of~10~mas corresponds to a linear distance
of~7.5~\hbox{AU} ($1\,\mathrm{mas} = 0.75\,\mathrm{AU}$), implying a
distance of~750~pc to the gas, though differences of as much 50\% are
possible.

For subsequent analysis, we fit the 2005 epoch optical depth line cube
by gaussian components, using the profile of Figure~\ref{fig:profile}
as an guide to initial values for the component parameters.  For the
fitting, we focussed on the three significant components identified.
The fitting was done on a pixel-by-pixel basis, with an independent
three-component fit for each pixel.  Table~\ref{tab:profile}
summarizes results of the fits, \emph{averaged} over the face of the
the source.  Guided by the results of the fitting from the 2005 epoch,
a similar fitting was performed for the 1998 epoch.

\begin{deluxetable}{lcccc}
\tablecaption{Optical Depth Profile Gaussian Component Fit
	Results\label{tab:profile}}
\tablewidth{0pc}
\tablehead{%
 \colhead{Component} & \colhead{Epoch}
	& \colhead{Central Velocity} 
	& \colhead{Velocity Width} & \colhead{Maximum Optical Depth} \\
                     &
	& \colhead{(\kms)} 
	& \colhead{(\kms)}         &
}
\startdata
1 & 1998 &     0.4 & 5.1 & 0.5 \\
  & 2005 &     0.3 & 4.9 & 0.5 \\
2 & 1998 &  $-8.0$ & 1.6 & 0.8 \\
  & 2005 &  $-8.0$ & 1.5 & 0.8 \\
3 & 1998 & $-10.4$ & 6.1 & 0.3 \\
  & 2005 & $-10.4$ & 6.2 & 0.3 \\
\enddata
\end{deluxetable}

\subsection{Small-Scale Structure}\label{sec:small}

\begin{figure*}
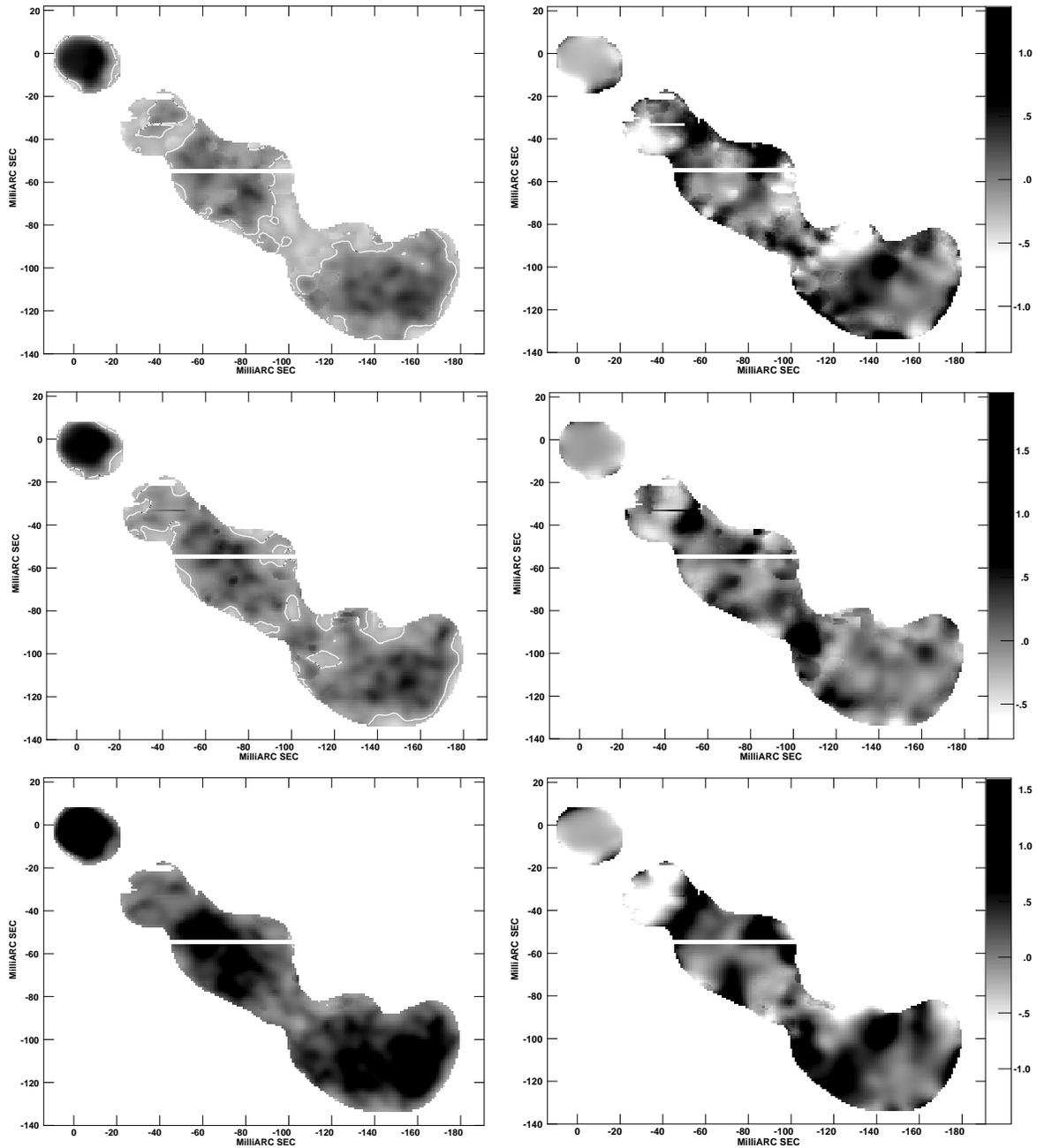

\begin{center}
\includegraphics[width=0.35\textwidth,angle=-90]{f3a.ps}~\includegraphics[width=0.35\textwidth,angle=-90]{f3b.ps}\\
\includegraphics[width=0.35\textwidth,angle=-90]{f3c.ps}~\includegraphics[width=0.35\textwidth,angle=-90]{f3d.ps}\\
\includegraphics[width=0.35\textwidth,angle=-90]{f3e.ps}~\includegraphics[width=0.35\textwidth,angle=-90]{f3f.ps}
\end{center}
\vspace*{-3ex}
\caption{%
Column density \emph{fluctuation} images, for the 1998 epoch
(Epoch~I), with the average value of the 2005 epoch column density
subtracted, $\Delta N_{\mathrm{H\,I}}/T_s \propto \int dv\,(\tau -
\langle\tau_{2005}\rangle)$.  Horizontal white lines are pixels where
the gaussian fitting failed to converge.
\emph{Left} panels show the signal-to-noise ratio, on a linear scale
with the dynamic range restricted to 0--15 and with a single contour
showing a signal-to-noise ratio of~5.
\emph{Right} panels show the column density fluctuations, on a linear
scale.  The gray scale bar shows the column density fluctuation, in
units of $10^{19}$~cm${}^{-2}$.
(\textit{Top})~Column density fluctuations for~$-10.4\,\kms$ velocity
component in~1998, the 2005 average \emph{optical depth} that was
subtracted is $\langle\tau_{2005}\rangle = 0.31$;
(\textit{Middle})~Column density fluctuations for~$-8.0\,\kms$ velocity
component in~1998, $\langle\tau_{2005}\rangle = 0.93$; and
(\textit{Bottom})~Column density fluctuations for~$0.4\,\kms$ velocity
component in~1998, $\langle\tau_{2005}\rangle = 0.54$.
}
\label{fig:fluctuationI}
\end{figure*}

\begin{figure*}
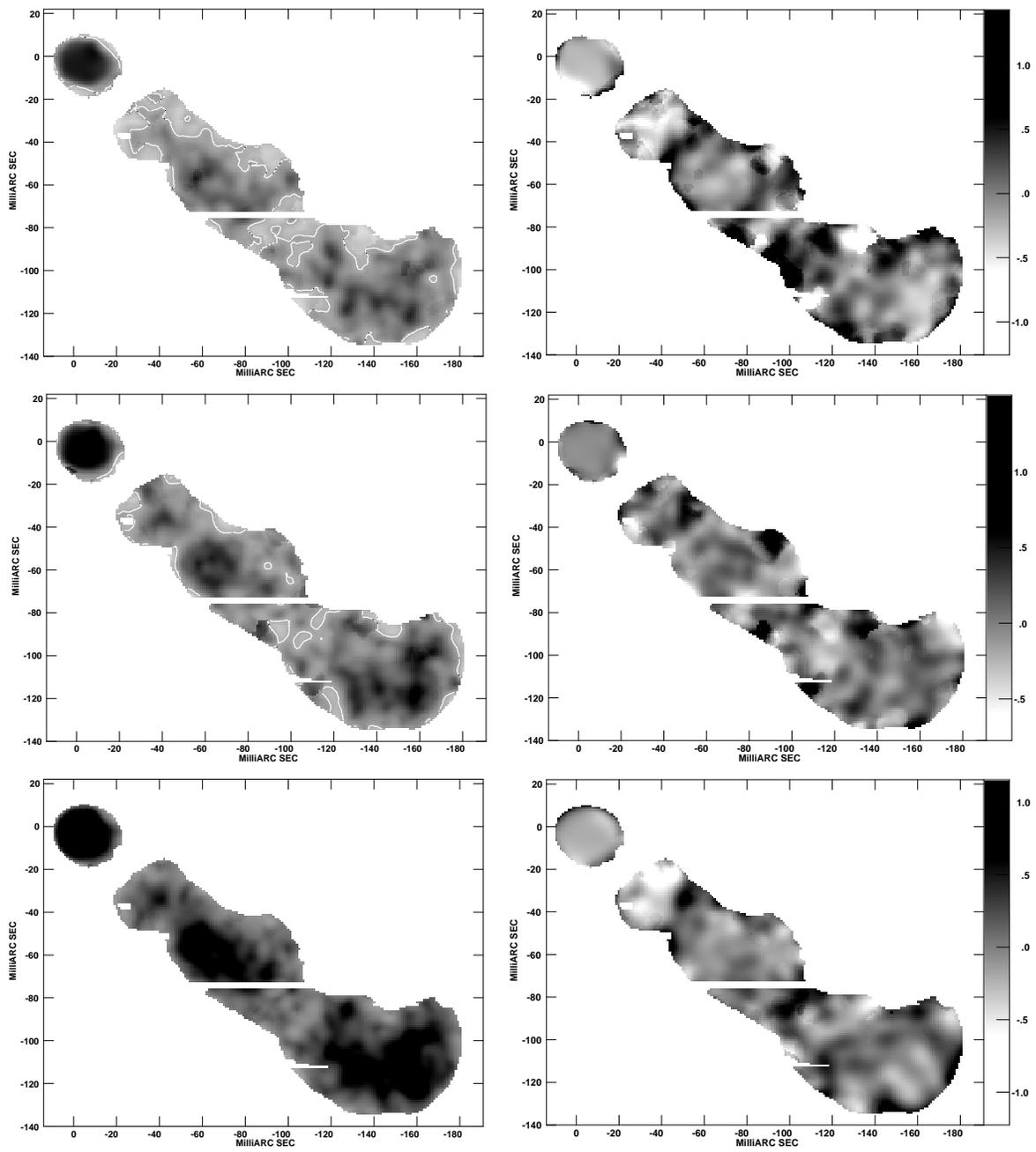

\begin{center}
\includegraphics[width=0.35\textwidth,angle=-90]{f4a.ps}~\includegraphics[width=0.35\textwidth,angle=-90]{f4b.ps}\\
\includegraphics[width=0.35\textwidth,angle=-90]{f4c.ps}~\includegraphics[width=0.35\textwidth,angle=-90]{f4d.ps}\\
\includegraphics[width=0.35\textwidth,angle=-90]{f4e.ps}~\includegraphics[width=0.35\textwidth,angle=-90]{f4f.ps}
\end{center}
\caption[]{Column density fluctuation images, as for
Figure~\ref{fig:fluctuationI} but for the 2005 epoch (Epoch~II).}
\label{fig:fluctuationII}
\end{figure*}

Figures~\ref{fig:fluctuationI} and~\ref{fig:fluctuationII} show \emph{column
density fluctuation} images at the two epochs for the three
different velocity components.  From the gaussian fits, column density
images were constructed from the fitting results by multiplying the
maximum optical depth by the velocity width ($N_{\mathrm{H\,I}}/T_s
\propto \tau\sigma_v$, see below regarding the spin
temperature~$T_s$).  In order to highlight fluctuations, the average
column density from the 2005 epoch across the face of the source was
subtracted from these column density images to produce the \emph{column
density fluctuation} images.  The signal-to-noise is not uniform across
the face of the source, and tends to decrease near the edges.  In
order to aid in assessing the reality of features,
Figures~\ref{fig:fluctuationI} and~\ref{fig:fluctuationII} also show
the column density fluctuation signal-to-noise ratio images.

We show column density fluctuation images, in contrast to
\cite{bzlgdf05} who showed optical depth channel images.  For \src,
analysis of the column density fluctuations is required because the
velocity field of the absorping gas appears to change somewhat within
each velocity component.  For \objectname[3C]{3C~138}, \cite{bzlgdf05}
compared the optical depth channel images at different velocities and
concluded that any velocity field fluctuations were negligible, in
contrast to the situation for \src.  Also,
in converting to column density, we assume a uniform spin temperature of the
gas of $T_s = 50$~\hbox{K} \citep{h97}.  Unlike
\objectname[3C]{3C~138} which  was part of the Millennium Arecibo
21~cm Survey \citep{ht03}, \src\ is outside the Arecibo declination
range so that there is less direct information about the absorbing gas
along this line of sight.

\begin{figure*}
\begin{center}
\includegraphics[width=0.47\textwidth,angle=-90]{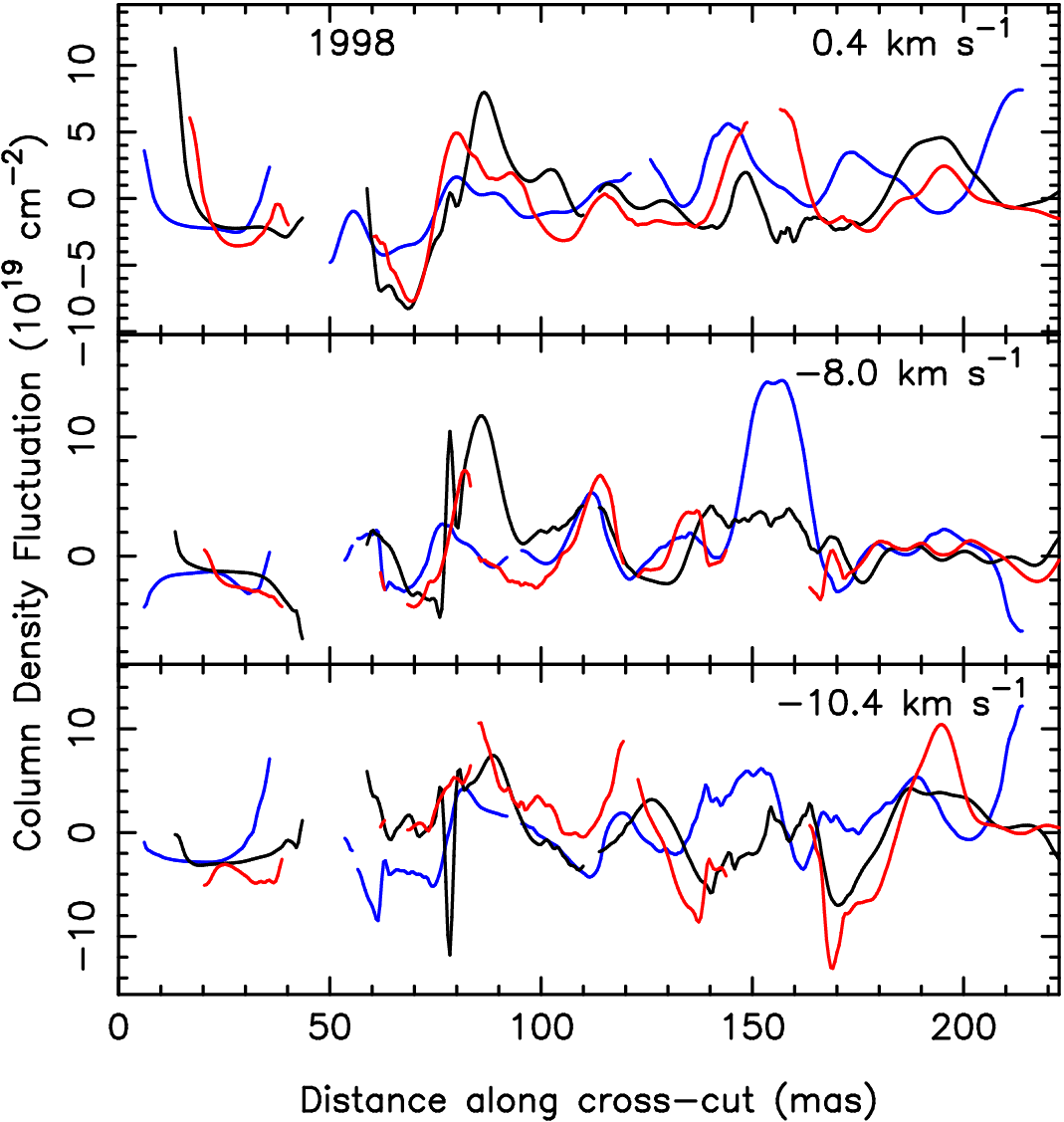}~\includegraphics[width=0.47\textwidth,angle=-90]{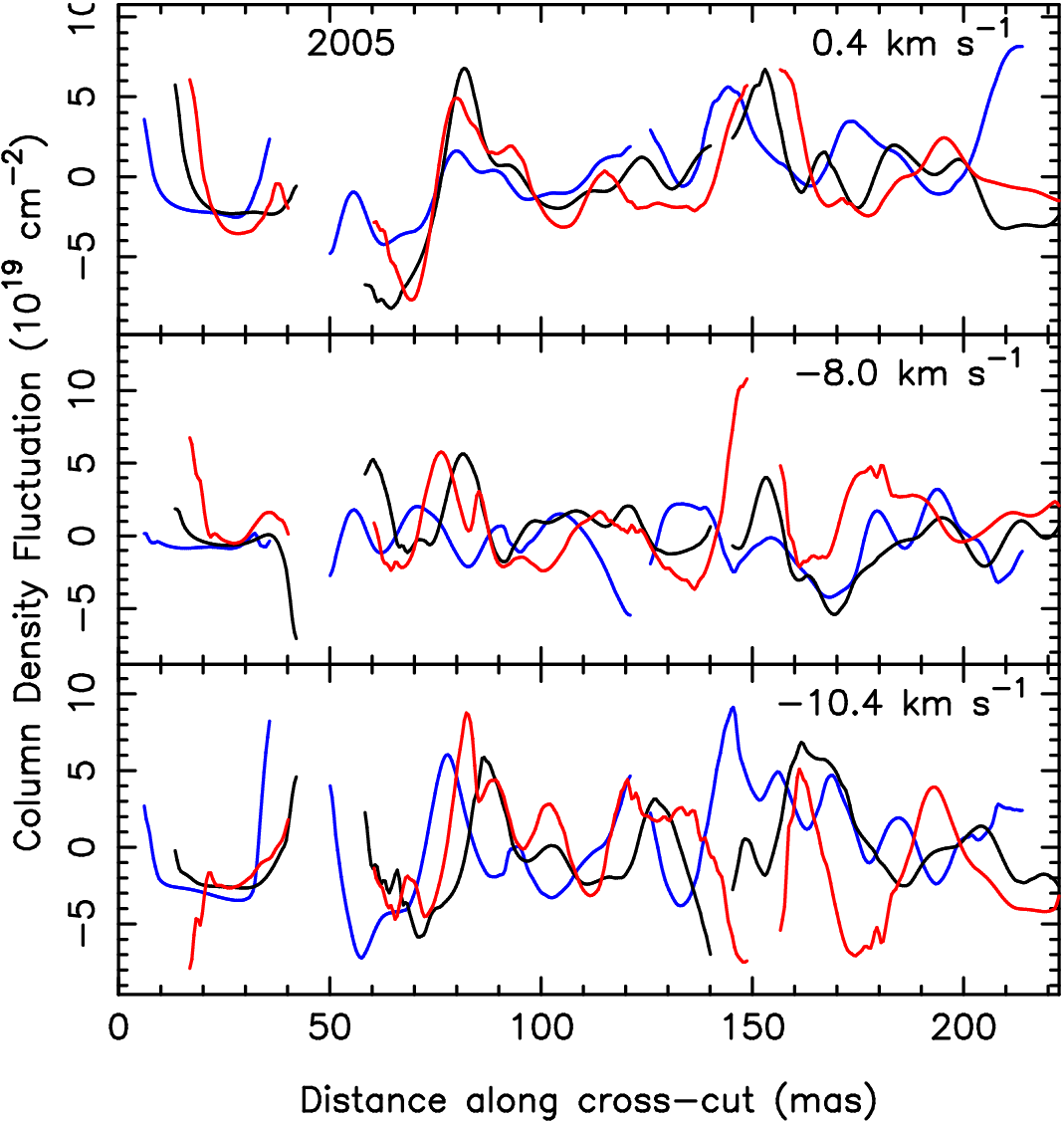}
\includegraphics[width=0.67\textwidth,angle=-90]{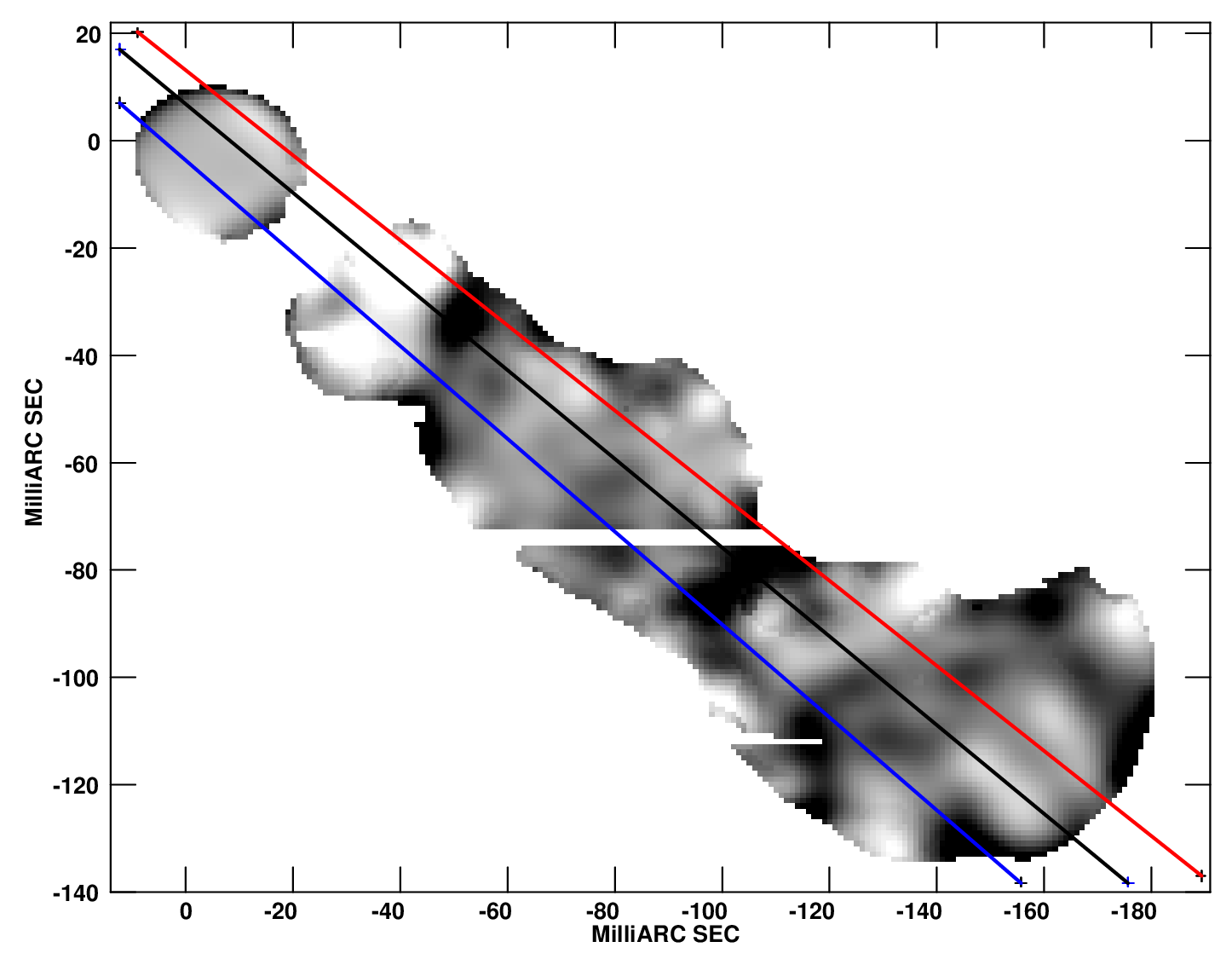}
\end{center}
\vspace*{-3ex}
\caption[]{
(\textit{Top Left}) Cross-cuts taken approximately along the major axis
showing the column density fluctuations for the 1998 epoch.
(\textit{Top Right}) Cross-cuts taken approximately along the major axis
showing the column density fluctuations for the 2005 epoch.
(\textit{Bottom}) Illustration showing where the cross cuts were taken
taken, with the column density fluctuations from the 0.4~\kms, 2005
epoch shown for reference.}
\label{fig:cross_major}
\end{figure*}

\begin{figure}[tb]
\begin{center}
\includegraphics[angle=-90,width=0.95\columnwidth]{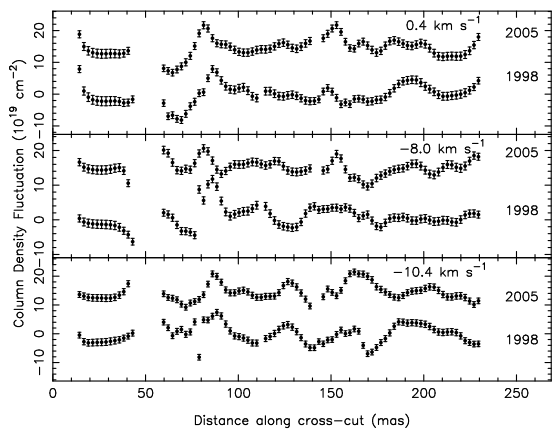}
\end{center}
\vspace*{-3ex}
\caption{Comparison of a cross-cut along the major axis at the two
epochs for the three velocity components.  For clarity, the 1998 epoch
is shifted (by~$15 \times 10^{19}$~cm${}^{-2}$) relative to the
2005 epoch.  For both epochs, a spin temperature $T_s = 50$~K is
assumed.  Also shown are uncertainties ($\pm 2\sigma$), although in
many cases they are only slightly larger than the symbol size.
Further, because the restoring beam can induce correlations, we plot
only every fifth datum.
The
major axis illustrated is the central of the three in
Figure~\ref{fig:cross_major}.}
\label{fig:slices_major}
\end{figure}

Figures~\ref{fig:cross_major}--\ref{fig:slices_minor} show cuts
through the column density fluctuation images.  In all cases,
spatially significant changes in the column density between the two
epochs are clearly apparent for all of the \ion{H}{1} components, at
significance levels exceeding $5\sigma$ over most of the face of the
source.  We have also conducted an analysis in which we consider a
constant column density cross-cut to be the null hypothesis.  In a
$\chi^2$ sense, the null hypothesis can be firmly rejected as typical
values, for both epochs and all velocity components, are $\chi^2 \sim
10$ (reduced $\chi^2$).  Typical angular scales for column density
variations are approximately 15~mas, corresponding to a linear scale
of approximately 10~\hbox{AU}.

\begin{figure*}
\begin{center}
\includegraphics[width=0.47\textwidth,angle=-90]{f6a.ps}~\includegraphics[width=0.47\textwidth,angle=-90]{f6b.ps}
\includegraphics[width=0.67\textwidth,angle=-90]{f6c.ps}
\end{center}
\vspace*{-3ex}
\caption[]{
As for Figure~\ref{fig:cross_major}, but for the minor axis.}
\label{fig:cross_minor}
\end{figure*}

\begin{figure}[tb]
\begin{center}
\includegraphics[angle=-90,width=0.95\columnwidth]{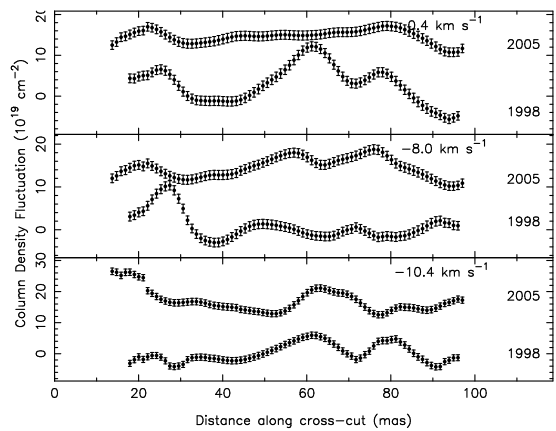}
\end{center}
\vspace*{-3ex}
\caption{As for Figure~\ref{fig:slices_major} (and
Figure~\ref{fig:cross_minor}), but for a minor axis slice.}
\label{fig:slices_minor}
\end{figure}

\cite{bzlgdf05} discussed the possible systematic effects that might
affect the extraction of reliable optical depth or column density
variations from dual-epoch imaging such as presented here.  We do not
repeat their discussion, but consider many of the same issues and
conclusions to hold.  Namely, while small differences in the images
from epoch to epoch may be due to the details of the observations and
data reduction (e.g., spatial frequency or $u$-$v$ coverage, slight
differences in the imaging), a number of steps were taken during the
analysis in an effort to minimize the differences between the epochs,
and the column density variations are significant.

The column density fluctuations in
Figures~\ref{fig:cross_major}--\ref{fig:slices_major} correpond to
peak-to-peak optical depth variations as large as $\Delta\tau \approx
0.7$, and typical optical depth variations on scales of approximately
15~mas ranges from 0.1--0.3.  The associated uncertainties in the
optical depth are $\sigma_\tau \approx 0.07$, implying significant
variations at the 3$\sigma$ level, and the column density cross-cuts
indicate variations at even higher significance are present.  Further,
the magnitude of the variations is approximately correlated with the
strength of the average \ion{H}{1} absorption toward \src, in that the
largest variations are observed at~$-8.0$~\kms, followed by 0.4~\kms, with
the smallest variations at~$-10.4$~\kms\ (cf.\
Figures~\ref{fig:profile} and
\ref{fig:cross_major}--\ref{fig:slices_minor}).

The opacity variations toward \objectname[3C]{3C~138} show changes
that are consistent with motion of structures across the line of sight
\citep{bzlgdf05}, though there is considerable uncertainty with making
these identifications (as they discuss).  Possible motions of
structures across the line of sight toward \src\ are also visible in
the cross-cuts.  Examples of such motions include the features at
distances between~50 and~100~mas (Figures~\ref{fig:slices_major}).
Caution in interpreting these features as arising from motions is
clearly warranted, however, given that we have only two epochs.
Nonetheless, typical position shifts appear to be of the order
of~5~mas.  Over the 7~yr interval between observations, the implied
proper motion is just under 1~mas~yr${}^{-1}$, equivalent to a
velocity of order 3~\kms, at a distance of~750~pc.  For comparison,
and recognizing that there is considerable uncertainty in these
comparisons, \cite{bzlgdf05} find larger values for the apparent
velocities of structures toward
\objectname[3C]{3C~138} ($\approx 20\,\kms$).

One of the motivations for undertaking these observations was to
assess whether the optical depth variations, both spatial and
temporal, found by \cite{bzlgdf05} toward \objectname[3C]{3C~138}
indicated that the line of sight to that source was in some sense
``special'' or anomalous.  Comparison of
Figures~\ref{fig:fluctuationI} and~\ref{fig:fluctuationII} and
Figures~\ref{fig:cross_major}--\ref{fig:slices_minor} with the
corresponding ones from \cite{bzlgdf05} show that they are
qualitatively similar, with clearly significant opacity or column
density variations occurring both in space and time.

Quantitatively, there are modest differences in the opacity/column
density variations between \src\ and \objectname[3C]{3C~138}.  We
estimate that the typical angular scale of opacity variations is
15~mas ($\approx 10$~AU), as opposed to about~50~mas ($\approx 25$~AU)
toward \objectname[3C]{3C~138}, though the linear scales are
comparable.  The magnitude of the variations toward \src\ also seems
somewhat smaller.  Optical depth changes (both in space and time), and
corresponding column density fluctuations, are a factor of a few to
several larger for \objectname[3C]{3C~138}---optical depth changes
of~0.4 and larger ($> 10^{20}$~cm${}^{-2}$) for
\objectname[3C]{3C~138} 
 vs.\ optical depths
typically not exceeding 0.3 ($\sim 5 \times 10^{19}$~cm${}^{-2}$) for \src.

\subsection{Small-Scale \ion{H}{1} Covering and Filling
	Factors}\label{sec:fill}

\cite{bzlgdf05} used their observations of \objectname[3C]{3C~138} to
conclude that the (two-dimensional) \emph{covering factor} of
small-scale \ion{H}{1} opacity variations was about~10\%, from which
they inferred a three-dimensional filling factor of probably less than
1\%.  Although the optical depth variations appear qualitatively
similar for lines of sight toward \objectname[3C]{3C~138} and \src, we
have repeated their analysis in order to determine the covering and
filling factors for the line of sight to \src.

\begin{figure}[tb]
\begin{center}
\includegraphics[angle=-90,width=0.95\columnwidth]{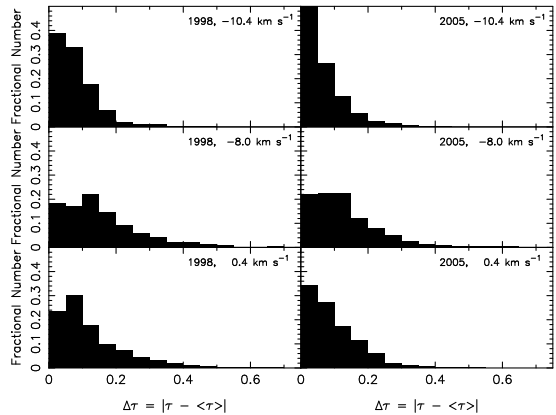}
\end{center}
\vspace*{-3ex}
\caption[]{Fractional number of pixels in an optical depth channel
image as a function of the optical depth variation $\Delta\tau \equiv
|\tau - \langle\tau\rangle|$.
\textit{Left} panels show the first epoch (1998), and
\textit{right} panels show the second epoch (2005).  
(\textit{Top})~$-10.4\,\kms$; 
(\textit{Middle})~$-8.0\,\kms$; and
(\textit{Bottom})~$0.4\,\kms$.}
\label{fig:fill}
\end{figure}

Figure~\ref{fig:fill} shows the fractional number of pixels in a
optical depth channel image for the three different velocities at both
epochs.  Most of the opacity variations are at a level less than
about~0.2 in optical depth, and we do not consider them significant.
Restricting to optical depth variations larger than approximately 0.2
($\approx 3\sigma$), most of the covering fractions are about~10\%,
ranging from a low value of a few percent
to a high exceeding 25\%.  

While we have decomposed the optical depth profile into gaussian
components, our decomposition is likely not unique and we fit only for
three components, whereas Figure~\ref{fig:profile} clearly shows that
there are could be more components.  Thus, a plausible upper limit to
the volume filling factor of the small-scale absorbing gas is obtained
by assuming that there are multiple components, each of which
contributes equally.  We obtain an upper limit of~1\%, a value which,
as \cite{bzlgdf05} emphasize, is not directly measurable.

\section{Discussion and Conclusions}\label{sec:conclude}

We have presented two epochs of observations of the \ion{H}{1} optical
depth across the face of the source \src\ on scales of approximately
10~mas.  The motivation for these observations was assessing whether
spatial and temporal \ion{H}{1} opacity variations found in
multi-epoch observations of \objectname[3C]{3C~138} by \cite{bzlgdf05}
were in some sense ``special'' or anomalous.

We find qualitatively similar opacity and column density variations
toward \src\ as were found toward \objectname[3C]{3C~138}.
Quantitatively, the variations toward \src\ appear to be somewhat
smaller in angular scale (15~mas vs.\ 50~mas) and smaller in magnitude
(by a factor $\sim 5$).  While the typical angular scale of the
variations toward \src\ appears smaller, the absorbing gas may be more
distant than that causing the absorption toward
\objectname[3C]{3C~138} (\S\ref{sec:observe}).  If so, the resulting
linear scales are comparable (10~AU for \src\ vs.\ 25~AU for
\objectname[3C]{3C~138}), though the uncertainties are large.

Further similarities are observed in the covering and filling factors
of the small-scale absorbing gas toward both sources.  For both lines
of sight, the covering factor appears to be approximately 10\%, and
the volume filling factor, while not measured directly, has a
plausible upper limit of~1\% (and potentially much less).

Both \objectname[]{3C~138} and \src\ display significant \ion{H}{1}
opacity variations across their faces, implying variations within the
ISM on scales of about~10 to~50~\hbox{AU} over path lengths ranging
from~100 to~1000~pc.  Using \hbox{MERLIN}, \cite{grmt08} also have
resolved significant \ion{H}{1} opacity variations across the faces of
\objectname[]{3C~111}, \objectname[]{3C~123},
and~\objectname[]{3C~161}, implying structure on scales of~50
to~500~\hbox{AU}.  We conclude that the conditions that cause such
small-scale variations are fairly widespread within the Galactic
\hbox{ISM}.  The reason that so few other sources display such
small-scale opacity variations is likely to be, as \cite{bzlgdf05}
discuss, that few sources other than \objectname[3C]{3C~138} and \src\
have the combination of angular extent and surface brightness required
to conduct these observations.  One unfortunate implication of this
conclusion is that milliarcsecond-scale \ion{H}{1} observations of
other sources will largely not be useful for probing the small-scale
structure without a significant increase in sensitivity.  One possible
target, particularly with the existing High Sensitivity
Array (HSA),\footnote{
The VLBA combined with other large aperture telescopes such as the
phased \hbox{VLA}, the Green Bank Telescope, Arecibo, or the 100-m
Efflesberg telescope.}
may be \objectname[3C]{3C~380}.

In one aspect, however, the lines of sight to \src\ and
\objectname[3C]{3C~138} do differ.  For the \objectname[3C]{3C~138}
analysis, \cite{bzlgdf05} used optical depth velocity channel images
whereas, for \src, we fit the optical depth line cube with gaussian
components and used the resulting column density images.  This
difference in approach was motivated by the velocity structure that
was apparent within an \ion{H}{1} component in the \src\ optical depth
line cube. 

We have not been able to find a ready explanation for this
difference.  Both sources are seen toward the Galactic anticenter,
with Galactic coordinates (longitude, latitude) of 
(161\fdg7, 10\fdg3) for \src\ and  (187\fdg4, $-11\fdg3$) for
\objectname[3C]{3C~138}, respectively.  To first order, the lines of
sight to both cut almost perpendicular to the Perseus spiral arm.
Further, were velocity crowding the explanation, it would seem that
that should be more of an issue for \objectname[3C]{3C~138} than for
\src.  We have also consulted the WHAM H$\alpha$ survey and the Green
supernova remnant (SNR) catalog \citep{g06}, reasoning that H$\alpha$ and SNRs
might serve as a tracers of turbulence injected by winds or explosions
from massive stars.  There are no obvious indications that the line of
sight to \src\ should be affected any such turbulence---indeed a
comparison of the \objectname[3C]{3C~138} and \src\ lines of sight
suggest that the line of sight to \objectname[3C]{3C~138} would be
\emph{more} likely to be the one that would display any such evidence
of turbulence.  

An alternate possibility is that the kinematic differences  between
these two lines of sight reflect small-scale features, and possibly
the past history of the gas.  
\cite{ksg85} imaged the \ion{H}{1} emission around the line of sight
toward \src\ at~1\arcmin\ resolution ($\approx 1$~pc linear scale).
They found a series of filaments and small clumps of \ion{H}{1}
emission, and they were able to associate at least some of the
absorption features with small emission clumps.  The amount of
small-scale structure (in \ion{H}{1} emission) toward \src\ is not
generally observed in the on-going GALFA \ion{H}{1} survey at Arecibo.
Further, \cite{ksg85} find that a large fraction ($\sim 80$\%) of the
\ion{H}{1} in emission in the direction of \src\ has a temperature
of~500--2000~\hbox{K}.  At this temperature, the gas would be
thermally unstable.  While the warm \ion{H}{1} is not responsible for
the absorption, the possibility that this line of sight contains
thermally unstable \ion{H}{1} is consistent with a scenario in which
the microphysics, and potentially the past history of the gas, leads
to kinematic variations within an \ion{H}{1} absorption component.

\cite{ksg85} also suggested a relative distance ordering of the gas.
Comparison of the
opacity variations (Figures~\ref{fig:cross_major}--\ref{fig:slices_minor}) suggests that
the opacity variations are smallest in amplitude for the $-10.4$~\kms\
velocity component and increase in magnitude for the 0.4~\kms\ and the
$-8.0$~\kms\ velocity components.  One interpretation is that the
opacity variations result from structures of essentially constant
size, which are comparable to or smaller than the equivalent linear
size of our beam ($\sim 10$~AU).  If this were the case, we could
obtain a relative distance ordering of the gas, with the $-8.0$~\kms\
material being the nearest, followed by the 0.4~\kms\ material, and
the $-10.4$~\kms\ material being the most distant.  An alternate
interpretation (see above) would attribute these differences to the
history of exposure of the gas to shocks or other interstellar
disturbances.  Whichever is the case, there is clearly significant
structure on large scales ($\sim 1$~pc), suggesting that such
structure could persist to smaller scales.  The combination of VLBA
and VLA data (as well as potentially MERLIN data) might be able to
explore the connection between the small- and large-scale opacity
variations.

\cite{bftsgh83} have set an upper limit (3$\sigma$) on the magnetic field
toward \src\ of $B_\parallel < 50\,\mu\mathrm{G}$, based on Zeeman
effect measurements in \ion{H}{1} spectra.  Under the standard
assumption that discrete \ion{H}{1} structures require densities $n
\sim 10^5$~cm${}^{-3}$ \citep{h97}, with a typical velocity width of
$v \approx 3\,\kms$ (Figure~\ref{fig:profile} and
Table~\ref{tab:profile}), one concludes that magnetic and turbulent
equipartition requires a magnetic field strength of order
\hbox{400~$\mu$G}, well above the observed upper limit.  As for the
optical depth variations toward 3C~138 \citep{bzlgdf05}, the optical
depth variations cannot be in magnetic and turbulent equilibrium,
unless there is significant blending and dilution of the magnetic
field on the angular scales over which the Zeeman effect measurements
were made.

Our observations of \src\ do not produce any new constraints on the
nature of the small-scale structures vis-a-vis whether they represent
``statistical'' fluctuations \citep[e.g.,][]{d00}, ``non-equilibrium''
physical structures \citep[e.g.,][]{jt01,ha07}, or discrete ``tiny
scale atomic structures'' \citep{h97}.  While the level of opacity
variations toward \src\ are lower than those toward
\objectname[3C]{3C~138}, we believe that this lower level can be
accommodated easily within any of these scenarios.  A wide range of
opacity variations might be expected if these result from
non-equilibrium processes, particularly because the level of opacity
variations could depend upon the history of the gas.  Also, as
\cite{bzlgdf05} note, the predicted level of opacity variations within
a statistical description depends sensitively upon the assumed
spectral index of the underlying power law; within the current
uncertainties for this spectral index, a large range of opacity
variations is allowed.

What would be required in order to place significant constraints on
these small-scale opacity variations?  Ideally, one should monitor the
same volume of gas and determine how the structures evolve.  In the
most simple comparison, discrete structures should show only 
linear motion across the line of sight, while fluctuations or
non-equilibrium physical conditions might also cause the appearance of
the opacity variations to change significantly.  

\begin{figure}[bt]
\begin{center}
\includegraphics[angle=-90,width=0.95\columnwidth]{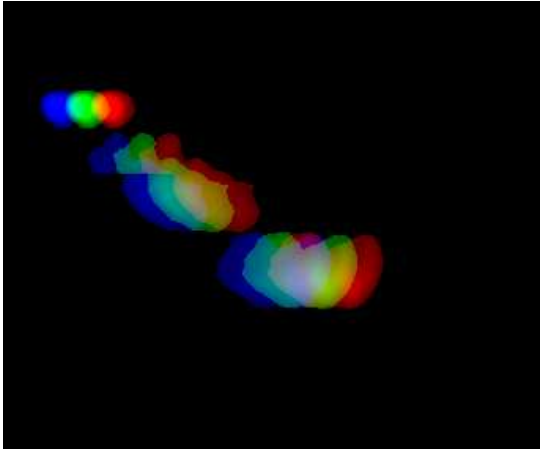}
\end{center}
\vspace*{-5ex}
\caption[]{An illustration of the different volumes of the Galaxy
sampled by multi-epoch VLBI observations.  The distance to the
absorbing gas is assumed to be 750~pc, the space velocity of the Sun
is assumed to be 30~\kms, and the gas is assumed to be stationary.
The resulting effective proper motion is 8.4~mas~yr${}^{-1}$; a
smaller assumed distance for the gas would result in a larger proper
motion while a smaller space velocity for the Sun would result in a
smaller proper motion.  Shown is the apparent position of the source
at three hypothetical epochs, each separated by~3~yr, comparable to
the typical separation in epochs for the existing multi-epoch VLBI
\ion{H}{1} absorption observations.  (Epoch~I is red, Epoch~II is
green, and Epoch~III is blue.)  The white areas indicate the
\emph{only} sampled volumes of gas common to all three epochs.}
\label{fig:cartoon}
\end{figure}

The elapsed times between observations presented here and those in
\cite{bzlgdf05} range from~3 to~7~yr.  These intervals are only a few
percent of the estimated time for discrete structures to change
substantially \citep[$\approx 500$~yr,][]{h07}.  However, on
milliarcsecond scales, the line of sight to \src\ effectively samples
a volume through the Galaxy \citep{mmb93,c08}.  The Sun's velocity
through space causes this volume to move between our observing epochs
(Figure~\ref{fig:cartoon}), in addition to any motion that the gas
itself might have.  A simple estimate of the Sun's motion suggests
that an entirely new volume through the Galaxy could be sampled on
time scales of approximately 3~yr.  That is, in the typical interval
between VLBI observations, essentially an entirely new volume of the
Galaxy is sampled by the line of sight.

Consequently, we also conclude that the time sampling of the existing
multi-epoch VLBI observations has been too coarse to distinguish
between the various models for the small-scale opacity variations.
Ideally, one would like to monitor the same volume of gas, to
determine if the opacity variations appear to be simply in motion or
also changing in appearance.  We estimate that, for either
\objectname[]{3C~138} and \src, an appropriate sampling interval is no
longer than about~9~months, with even more rapid sampling desirable.

\acknowledgements
We thank N.~Dieter-Conklin for helpful discussions on the motion of the lines
of sight through the Galaxy.
We thank J.~Dickey, the referee, who made insightful comments that we
believe improved the analysis presented here.
The Wisconsin H-Alpha Mapper is funded by the National Science
Foundation.
This research has made use of the SIMBAD database, operated at
\hbox{CDS}, Strasbourg, France.
This research has made use of NASA's Astrophysics Data System.
The National Radio Astronomy Observatory is a facility of the National
Science Foundation operated under cooperative agreement by Associated
Universities, Inc.  Basic research in radio astronomy at the NRL is
supported by 6.1 NRL Base funding.

\end{document}